# NANOCHEMISTRY OF FULLERENE $C_{60}$. CYANO- AND AZO-POLYDERIVATIVES

E.F.Sheka

Peoples' Friendship University of Russia, 117197 Moscow, Russia
sheka@icp.ac.ru

**ABSTRACT.** Cyanation $C_{60}$ to $C_{60}(CN)_{18}$ and aziridination from $C_{60}$ to $C_{60}(NH)_9$ have been studied by unrestricted broken spin symmetry Hartree-Fock approach implemented in semiempirical codes based on AM1 technique. The calculations were focused on successive addition of CN and NH moieties to the fullerene cage following the indication of the cage target atoms by the highest atomic chemical susceptibility calculated at each step. The obtained results are analyzed from the viewpoint of criteria on parallelism between these derivatives as well as $C_{60}$ fluorides and hydrides. The difference of the first stage $C_{60}$ chlorination from other sterically free processes is discussed.

**KEYWORDS**: cyano- and hydrocyano $C_{60}$ fullerenes; $C_{60}$-fulleroaziridines, $C_{60}$ chlorides, quantum chemistry; unrestricted broken symmetry approach; atom chemical susceptibility; computational synthesis

## 1. INTRODUCTION

This study has been undertaken to answer question how general are arguments favoring electronic and structural parallelism of $C_{60}$ fluorides and hydrides disclosed both experimentally and computationally and described in details in [1]. Actually, besides their individual intriguing peculiarities, two features are common for both fluorinated and hydrogenated fullerenes $C_{60}$. The former is evidently connected with the absence of sterical limitations, which is a result of a monoatomic structure of the addends and their small size. The latter concerns 1,2-fashion of the atom additions to the fullerene cage. A natural question arises how common are the two criteria for the structural parallelism to be observed in other cases when the requirements are met. From the computational viewpoint, a crown-like $C_{3v}$ structure for the 18-fold members of the family of polyderivatives seemed to be the best manifestation of the serial event. It was hard to overcome the temptation to look for other 18-fold derivatives which might be expected to fit two requirements mentioned above and to look at their structure. CN and NH units as possible addends look like the best candidates. Actually, their addition to the $C_{60}$ cage is not accompanied by strong steric limitations, on one hand, and must obey to the 1,2-addition algorithm, on the other. Just performed calculations have shown that the structural parallelism for both sets of derivatives with fluorinated and hydrogenated fullerenes $C_{60}$ families takes place indeed. But when we turned our attention upon chlorine, it turned out that the empirical reality definitely told 'no' to any parallelism of chlorides and, say, fluorides. On the first glance, the strictness of the above two criteria seems not be violated drastically. Is there any extra reason to provide parallelism of $C_{60}$ polyderivatives? The current paper suggests a wished answer.

## 2. GROUNDS OF COMPUTATIONAL METHODOLOGY

As in the papers devoted to the computational synthesis of $C_{60}$ fluorides [2] and hydrides [3], the algorithmic approach, based on precalculated atomic chemical susceptibility (ACS), is explored to computationally synthesize two families of $C_{60}$ polyderivatives related to $C_{60}(CN)_n$ and $C_{60}(NH)_m$ compounds. The stepwise successive synthesis is limited to $n=18$ and $m=9$ since linear

CN units are added to individual carbon atoms of the fullerene cage while every NH unit blokes one C-C bond due to interaction of the nitrogen atoms with two carbons. The limitation is subordinated to checking if the numbers $n=18$ and $m=9$ might be 'magic' ones if suppose that the $C_{3v}$-crown structure of polyderivatives in this case is the brightest manifestation of the parallelism.

## 3. POLYHYDROCYANIDES $C_{60}H(CN)_{2n-1}$ AND POLYCYANIDES $C_{60}(CN)_{2n}$

Cyano- and hydrocyano $C_{60}$ fullerenes invented by Wudl et al. [4, 5] have attracted a great attention as components of proton exchange membrane fuel cells [6, 7] due to their high proton conductivity. The inventors have elaborated a multistep route of producing polyhydrocyanofullerenes from $C_{60}$ to $C_{60}H(CN)_5$ that can be evidently expanded over higher members of the family and presents a series

$$C_{60} \to C_{60}H(CN)_1 \to C_{60}(CN)_2 \to C_{60}H(CN)_3 \to C_{60}(CN)_4 \to \ldots \to C_{60}H(CN)_{2n-1} \to C_{60}(CN)_{2n} .$$

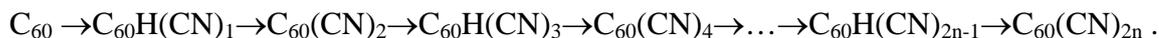

Series 1

This production route has been supported by other chemists (see for example [8, 9]) demonstrating the best way to produce hydrocyano $C_{60}H(CN)_{2n-1}$ and cyano $C_{60}(CN)_{2n}$ polyderivatives and their ions. Different molecular and anionic representatives of the two series, which cover members with $n=1$, 2, and 3, have been obtained by now.

Let us let lay the foundation of the production algorithm of Series 1 into the background of stepwise computational synthesis of the species. Although in practice alkali metal cyanides are the main reagents when obtaining anion salts of fullerene $C_{60}$, quenched afterwards by different acids to get the products of the series [4], it is evident that in computational experiment the products formation can be considered as a result of intermolecular interaction between the partners of the $C_{60}$ + HCN dyad. Calculations were performed within the framework of the unrestricted broken symmetry Hartree-Fock approach by using AM1 version of its semiempirical implementation [10]. As it turns out, the final product depends on the starting distance between C and H atoms of the molecule with the cage atoms $C_c$. Particularly it is critical with respect to the hydrogen atom. If HCN is aligned along one of C-C bonds joining two carbon atoms ($C_c$) of fullerene belonging to group 1 with the highest ACS (see detailed description in [2, 3]), the molecule is added to the fullerene cage associatively (product **I**) (see Fig. 1) when the $C_c$-H distance exceeds 1.7Å. When the distance is less then 1.4Å, the molecule is attached to the fullerene cage dissociatively (product **II**). The energy gain constitutes 48,5 kcal/mol in favor of product **II** so that the formation of the first product of Series 1 is energetically profitable. In turn, the interval of 1.4-1.7Å marks the space where the barrier of the HCN molecule dissociation in the vicinity of fullerene $C_{60}$ is positioned.

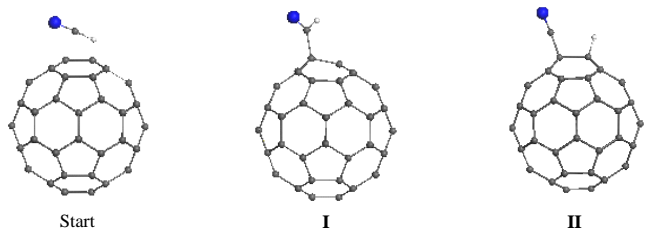

Start     **I**     **II**

**Figure 1**. The first meeting of HCN molecule with fullerene $C_{60}$. A general starting composition and equilibrium structures of products **I** and **II** (see text). Small black balls point carbon atoms while a small white and big blue balls mark hydrogen and nitrogen atoms, respectively. UBS HF AM1 singlet state.

Starting from product **II**, keeping its equilibrium structure and orientation of the hydrogen atom, let us substitute the hydrogen atom with the second CN unit and continue the structure optimization. Thus, the equilibrium structure of bis-cyanofullerene $C_{60}(CN)_2$ is obtained. Its ACS map is analyzed for two new carbon atoms to be selected for the next step of the HCN molecule dissociative addition. As it turns out, similarly to the case of fluorides and hydrides [2, 3], these two atoms are located in the equatorial space of the fullerene cage. The addition of HCN molecule to these atoms results in the $C_{60}H(CN)_3$ formation. Substituting the hydrogen atom of the species by CN unit, one obtains tetra-cyanofullerene $C_{60}(CN)_4$, the ACS map of which points to new carbon atoms for the next HCN addition resulting in the $C_{60}H(CH)_5$ production, and so forth. It should be noted that in contrast to the fluorides and hydrides case, the difference between the highest rank $N_{DA}$ and remaining data in the $N_{DA}$ list is well pronounced, so that there is no necessity in isomer analysis by energy in this case. Following this algorithm, the final product, $C_{60}(CN)_{18}$, of this series was obtained.

Figure 2 presents odd members of Series 1 related to the obtained hydrocyano[$C_{60}$] fullerenes. The structures are drawn in a fixed projection related to the fullerene cage to make possible a viewing of successive change in the structure of the molecules as a whole and the fullerene cage in particular. The even members of the series related to polycyanofullerenes are shown in Fig. 3. The molecules are presented in a particular projection - let us call it as central-hexagon one - to exhibit the evolution of the molecule as a whole and the fullerene cage structure towards that of 18CN of the exact $C_{3v}$ symmetry. The molecule is of a crown structure similar to those of $C_{60}F_{18}$ and $C_{60}H_{18}$. Two last panels in the right down corner compare crown structures of 18CN and 18F molecules from [2]. As seen from both figures, no vivid steric complications accompany the successive cyanation of the fullerene cage performed as a successive chain of 1,2-additions. The carbon skeleton preserves its closed shape and responds to the deformation caused by the cyanation just minimizing angular and bond strengths in a similar way that is characteristic for fluorination and hydrogenation.

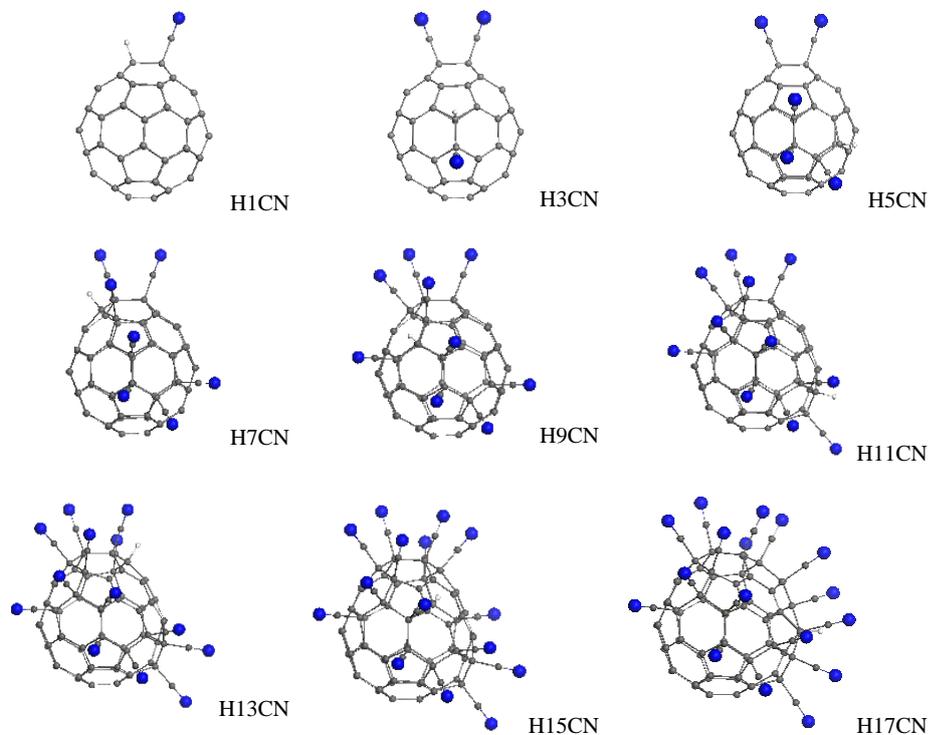

**Figure 2**. Equilibrium structures of hydrocyanofullerenes $C_{60}H(CN)_{2n-1}$ (n=1, …, 9). To simplify the product notation a concise marking H(2n-1)CN is used. Atoms marking see in caption to Fig. 1.

A successive change in the fullerene skeleton structure under cyanation is presented in Fig. 4 for a fixed set of C-C bonds. The main changing corresponds to the formation of long C-C bond between cyanated carbon atoms and neighboring atoms of non-cyanated skeleton. The bond length starting at 1.55Å for bis-cyanofullerene gradually arises and achieves 1.67Å for high members of the series. So large increasing in the bond length is similar to that observed for $C_{60}$ fluorides [2] and is a measure of the skeleton deformation caused by $sp^2$-$sp^3$ transformation of the carbon atom configuration.

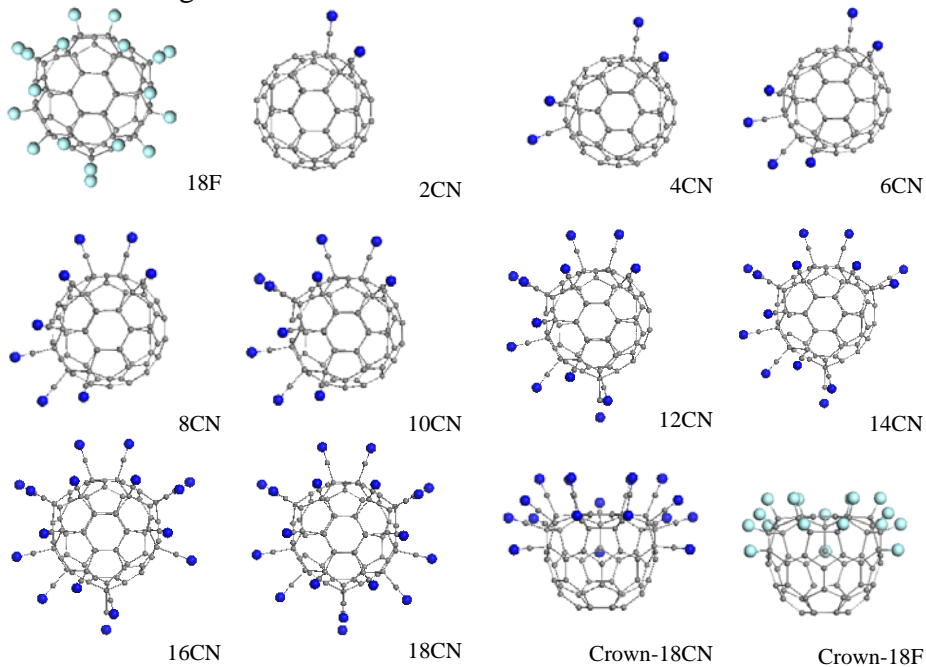

**Figure 3**. Equilibrium structures of cyanofullerenes $C_{60}(CN)_{2n}$ (n=1, …, 9) in the central-hexagon projection. To simplify the product notation a concise marking 2nCN is used. For atom' marking see caption to Fig. 1.

The product symmetries and total energies are listed in Table 1. According to two-different-stage production of the series, the coupling energy $E_{coupl}$ that is needed to produce a new species at each successive step should be estimated separately. Thus, the production of hydrocyanofullerenes is described by $E_{cpl}$, determined as

$$E_{cpl}[H(CN)_{2n-1}] = \Delta H[H(CN)_{2n-1}] - \Delta H[(CN)_{2n-2}] - \Delta H(HCN), \quad (1)$$

where the right-hand-part components present the total energies (heats of formation) of hydrocyanofullerene $C_{60}H(CH)_{2n-1}$, cyanofullerene $C_{60}(CN)_{2n-2}$, and HCN molecule, respectively. In its turn, the production of cyanofullerenes is described by the coupling energy $E_{cpl}$

$$E_{cpl}[(CN)_{2n}] = \Delta H[(CN)_{2n}] - \Delta H[H(CN)_{2n-1}] - \Delta H(CN). \quad (2)$$

Here the right-hand-part components present the total energies of cyanofullerene $C_{60}(CN)_{2n}$ hydrocyanofullerene $C_{60}H(CH)_{2n-1}$, and the CN molecular unit, respectively. The corresponding energy coupling data are listed in Table 6.1. As seen from the table, the serial production is energetically favorable. $C_s$ symmetry for compound H1CN and $C_{2v}$ symmetry for compound 2CN were assigned on the basis of $^1$H NMR, $^{13}$C NMR, UV-vis, and IR spectra [5]. The latter finding is fully consistent with calculations. As for the former, the calculations attribute the H1CN structure to $C_1$ symmetry. However, a detailed continuous symmetry analysis in the framework of the methodology discussed in [11] gives 100% $C_s$-ness for the $C_1$ H1CN structure thus removing the inconsistence.

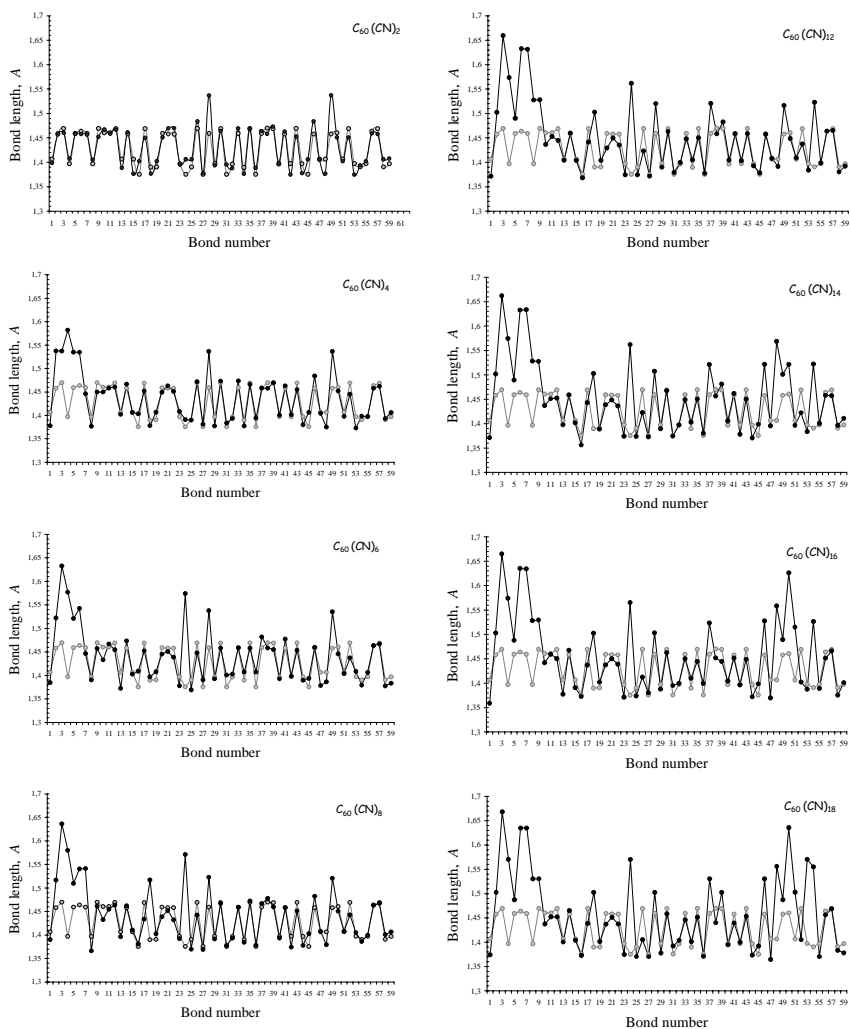

**Figure 4**. $sp^2$-$sp^3$ Transformation of the $C_{60}$ cage structure in due course of successive cyanation. Gray-dot and black-dot diagrams correspond to the bond length distribution related to the $C_{60}$ cage of pristine fullerene and its cyanated derivatives, respectively.

**Table 1**. Total energy and symmetry of polycyanides of fullerenes $C_{60}$

| | Polycyanides $C_{60}$ (CN)$_{2n}$ | | | | | | | | |
|---|---|---|---|---|---|---|---|---|---|
| | **2CN** | **4CN** | **6CN** | **8CN** | **10CN** | **12CN** | **14CN** | **16CN** | **18CN** |
| $\Delta H^1$, kcal/mol | 996.58 | 1042.44 | 1087.45 | 1146.94 | 1102.22 | 1247.62 | 1298.09 | 1349.46 | 1400.48 |
| $E_{cpl}$, kcal/mol | -67.39 | -60.76 | -65.40 | -64.02 | -66.41 | -64.20 | -67.39 | -60.23 | -90.36 |
| Symmetry | $C_{2v}$ | $C_s$ | $C_1$ | $C_s$ | $C_s$ | $C_1$ | $C_s$ | $C_1$ | $C_{3v}$ |
| | Polyhydrocyanides $C_{60}$ H(CN)$_{2n-1}$ | | | | | | | | |
| | **H1CN** | **H3CN** | **H5CN** | **H7CN** | **H9CN** | **H11CN** | **H13CN** | **H15CN** | **H17CN** |
| $\Delta H^1$, kcal/mol | 951.70 | 990.93 | 1040.59 | 1098.68 | 1148.24 | 1199.55 | 1246.63 | 1297.47 | 1383.57 |
| $E_{cpl}$, kcal/mol | -34.65 | -36.63 | -32.84 | -34.41 | -29.69 | -33.42 | -31.98 | -31.61 | -24.65 |
| Symmetry | $C_1$ | $C_s$ | $C_s$ | $C_1$ | $C_s$ | $C_1$ | $C_1$ | $C_1$ | $C_1$ |

[1] Molecular energies are presented by heats of formation $\Delta H$ determined as $\Delta H = E_{tot} - \sum_A (E_{elec}^A + EHEAT^A)$. Here $E_{tot} = E_{elec} + E_{nuc}$, while $E_{elec}$ and $E_{nuc}$ are the electron and core energies. $E_{elec}^A$ and $EHEAT^A$ are electron energy and heat of formation of an isolated atom, respectively.

## 4. POLYAZODERIVATIVES $C_{60}(NH)_m$

Fullerene $C_{60}$ derivatives, which involve nitrogen atoms within the attached units, present the largest class of chemicals produced on the fullerene basis by now (see monographs [12, 13]). The product structure critically depends on the addend atomic composition, particularly in the case of polyadditions. In the current paper we will consider a family of azoderivatives of fullerene $C_{60}$ [14] that are rooted in a large class of fulleroaziridines. A small, two-atom composition of the addend (NH) makes it interesting from the viewpoint of expected structural parallelism. On the other hand, since attaching an NH unit to the fullerene cage occurs bidentatly through the formation of two C-N bonds, 1.2-fashion of the additions is supported as well. The computational synthesis under discussion covers a series of polyfulleroaziridines $C_{60}(NH)_m$ for $m=1, 2,…, 9$. $C_{60}(NH)_9$ aziridine can be compared with corresponding 18-added-units fluoride, hydride, and cyanide.

Full set of synthesized products is shown in Fig. 5. As previously, the first product $C_{60}(NH)_1$ is obtained when attaching the nitrogen atom to a couple of skeleton carbon atoms of group 1 joined by a short bond. The $N_{DA}$ map of the $C_{60}(NH)_1$ product points to two equivalent pairs of carbon atoms situated in the equatorial region of the cage, which results in the production of the second aziridine $C_{60}(NH)_2$. The ACS map of the latter exhibits a couple of carbon atoms in the direct vicinity of the first addition thus causing the formation of aziridine $C_{60}(NH)_3$. Following this algorithm, the final product of the series $C_{60}(NH)_9$ was produced. Similarly to cyanides, the high rank $N_{DA}$ data are well separated from the others in the list, which is why the presented synthesis does not require an additional isomer analysis by energy and only monomer products should be expected in due course if a real synthesis.

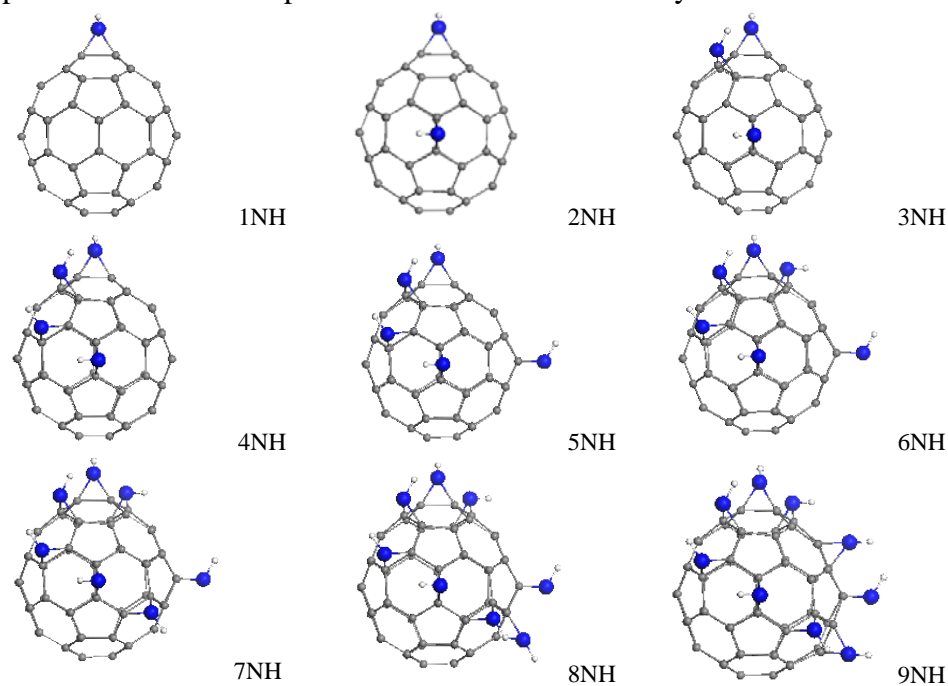

**Figure 5**. Equilibrium structures of polyaziridines $C_{60}(NH)_m$ (m=1, …, 9). To simplify the product notation a concise marking mNH is used. For atom' marking see caption to Fig. 1. UBS HF AM1 singlet state.

The structure projection in Fig. 5 is similar to that in Fig. 2, making it possible to compare both a general pattern and the difference in details under the successive additions in the two cases. As seen in the figure, the succession of additions in both cases is different, in spite of which the final structures appearing to be similar. Another view on the aziridine's structure is given in Fig. 6. As previously, the central-hexagon projection was chosen to trace the origin and development of the $C_{3v}$-symmetry pattern if it occurs. As seen in the figure, the nitrogen-carbon carcass of the $C_{60}(NH)_9$ aziridine has an evident characteristic $C_{3v}$-pattern. However, due to large dispersion of the N-H bond space orientations at practically the same energy, which causes a pronounced scattering in the hydrogen atom positions, the exact symmetry of the aziridine is $C_1$. However, the structure analysis from the viewpoint of continuous symmetry [11] shows that the structure is of 99.7% $C_{3v}$ so we can convincingly say about a deep structural parallelism of this aziridine with 18-fold attached fluorides, hydrides, and cyanides.

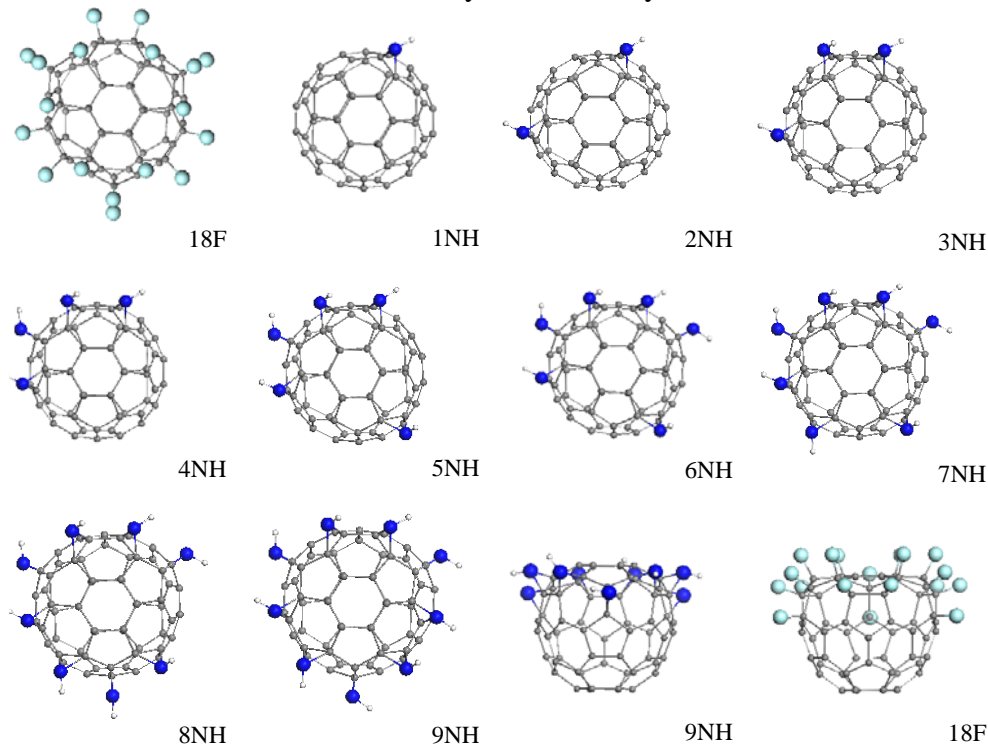

**Figure 6**. Equilibrium structures of polyaziridines $C_{60}H(CN)_m$ (m=1, …, 9) in the central-hexagon projection.

The additions of NH units to the fullerene cage, as of all other addends considered earlier, causes a noticeable deformation of the latter. Figure 7 presents a successive deformation of the cage in terms of selected C-C bonds throughout the aziridination. The product symmetries and total energies are listed in Table 2. The evolution of the total energy of the products and coupling energy needed for the every next addition of NH unit is presented in Fig. 8. The coupling energy is determined as

$$E_{coupl}[(NH)_m] = \Delta H[(NH)_m] - \Delta H[(NH)_{m-1}] - \Delta H(NH). \quad (3)$$

As seen in the figure, the total energy gradually grows when aziridination proceeds. In spite of the fact, the coupling energy, determined according to Eq.(3), remains negative and quite large by the value. This means that the serial production is energetically favorable. The evolution of the total number of effectively unpaired electrons in the famil, shown in the top panel of Fig. 8*b*, demonstrates high chemical reactivity of all members of the series.

So far there has been the only report [14] on the production and characterization of the first member of the series attributing its symmetry to $C_{2v}$ point group. Such occurred that this very substance (noted as molecule **II**) was discussed in [11] in regards to continuous symmetry of fullerene derivatives and their electronic spectra and had to make a conclusion about high continuous symmetry of the $C_1$ molecule consisting of ~94% $I_h$ symmetry, this was supported by high-symmetry pattern of its optical spectra close to $I_h$ one. Obviously, a high $I_h$ grade of continuous symmetry implies that symmetric patterns related not only to $I_h$ itself but to its subgroups might be expected. In this view, attributing high-symmetry pattern of $^{13}$C NMR spectra of the species to $C_{2v}$ group [14] does not seem strange and inconsistent with computational predictions. The continuous symmetry contribution of $C_{2v}$ into 9NH molecule structure constitutes 98.7%.

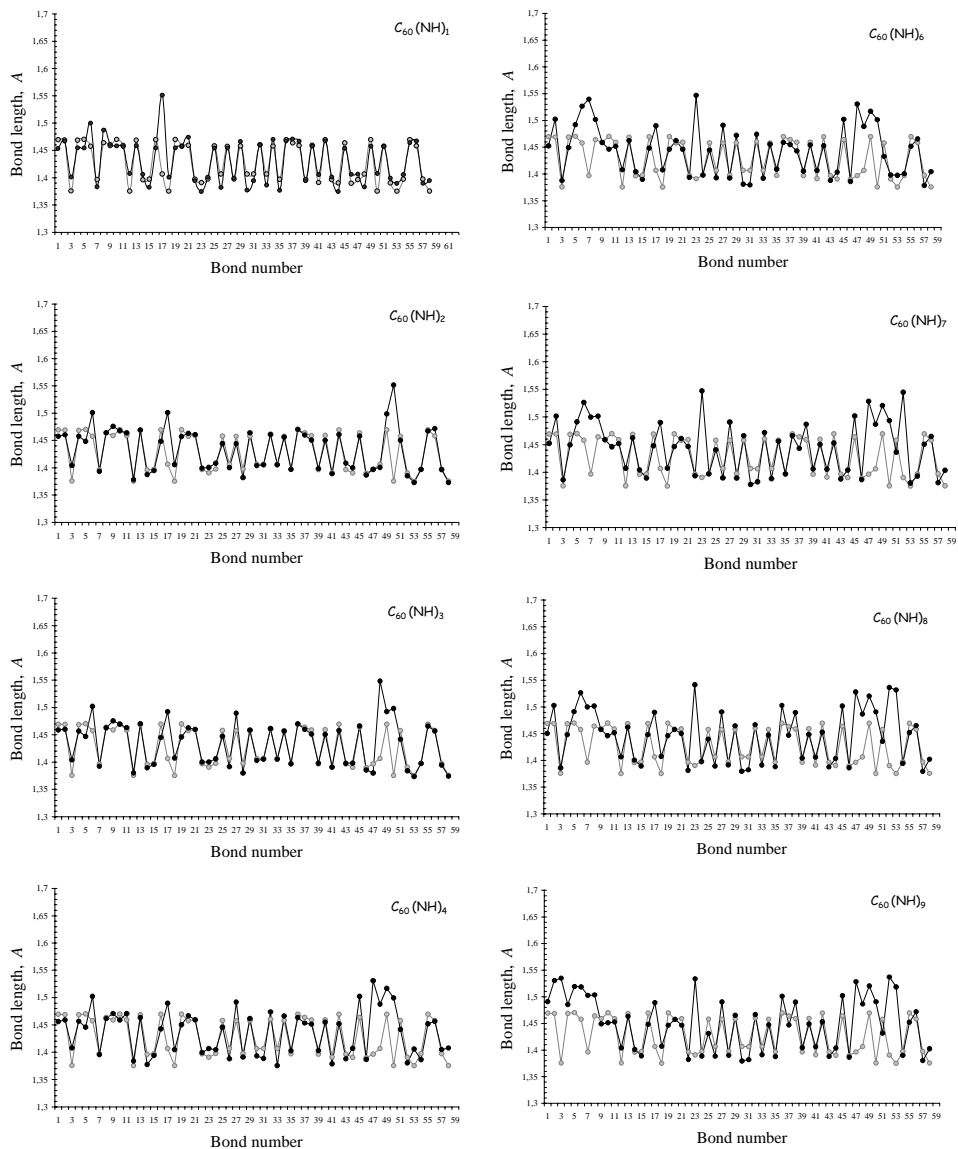

**Figure 7**. $sp^2$-$sp^3$ Transformation of the C$_{60}$ cage structure in due course of successive aziridination. Gray-dot and black-dot diagrams correspond to the bond length distribution related to the C$_{60}$ cage of pristine fullerene and its fulleroaziridines, respectively.

**Table 2**. Total energy and symmetry of parent polyazidines of fullerenes C$_{60}$

| | Polyazirinides C$_{60}$ (NH)$_m$ | | | | | | | | |
|---|---|---|---|---|---|---|---|---|---|
| | **1NH** | **2NH** | **3NH** | **4NH** | **5NH** | **6NH** | **7NH** | **8NH** | **9NH** |
| $\Delta H^1$ kcal/mol | 965.24 | 974.24 | 990.05 | 992.57 | 1006.21 | 1011.58 | 1030.02 | 1038.36 | 1051.78 |
| Symmetry | $C_1$ | $C_1$ | $C_1$ | $C_1$ | $C_1$ | $C_1$ | $C_1$ | $C_1$ | $C_1$ |

[1] See footnote to Table 1.

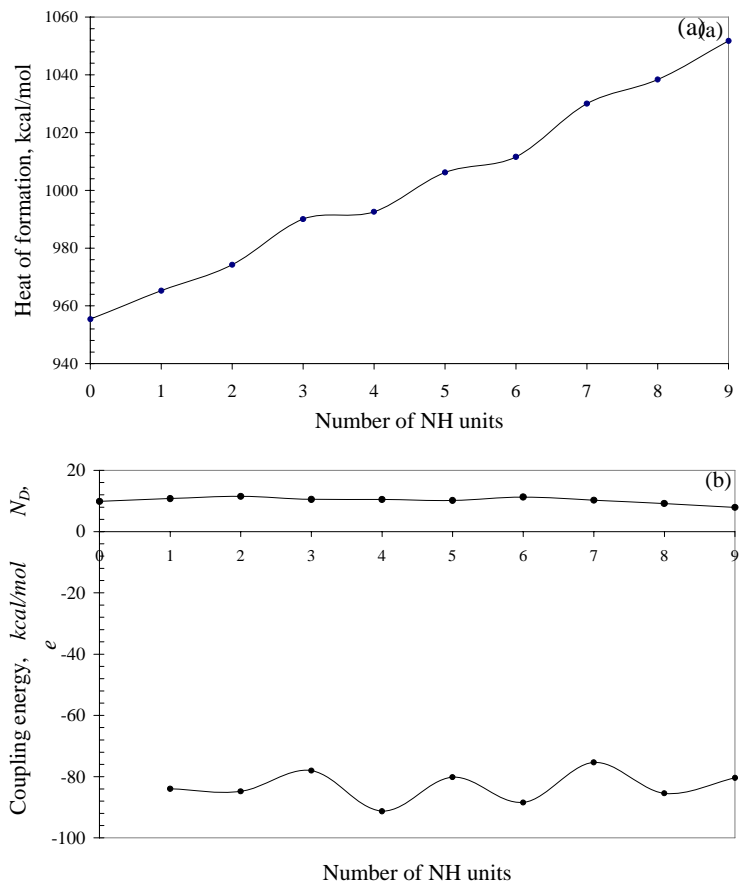

**Figure 8.** Evolution of the total energy (*a*), molecular chemical susceptibility, $N_D$, and coupling energy, $E_{cpl}$, (*b*) at growing the number of CN units attached to $C_{60}$ skeleton.

## 5. CONCLUSING REMARKS: A LITTLE ABOUT $C_{60}$ CHLORINATION

Our suggestions concerning a deep similarity in the structural patterns of the fullerene $C_{60}$ derivatives when the relevant additions occurs without sterical limitations and in the 1,2-fashion seem to become true. Additional support of the statement can be obtained when comparing structures of the 18-fold $C_{3v}$ crown members of all four families in Fig. 9. When looking at the picture one should keep in mind that the bond sets are the same only in the case of fluorination and hydrogenation, while other two sets are grouped in a different manner due to changing the atom numeration in the pristine fullerene molecule in two latter cases. Nevertheless, the presented picture reveals quite clearly both a big similarity in the general behavior of the bond distribution and the different extent of the cage deformation under different additions. Obviously, fluorination and cyanation cause the largest effect, while hydrogenation and aziridination disturb the cage to a much weaker but comparable extent.

Evidently the absence of sterical limitations and the 1,2-fashion of successive additions are necessary to support a deep parallelism of the relevant derivatives. However, a question arises if they are enough as well to cover all kinds of possible addition reactions that meet the requirements. Thus, we cannot pass by $C_{60}$ chlorination that occurs quite differently in practice in comparison with fluorination and hydrogenation [15]. Evidently, sterical limitations are more severe in the case. However, they are not strong enough to explain the drastic difference. We are not going to deeply discuss the chlorination chemistry of fullerene $C_{60}$ and will restrict ourselves by only a short excursus into first-step computational synthesis of $C_{60}$ chlorides within the framework of the approach based on ACS's guiding role.

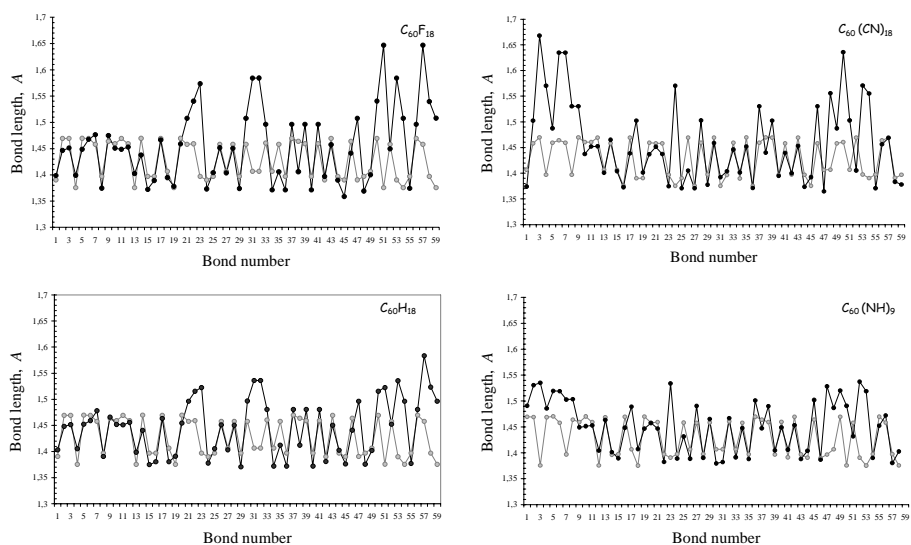

**Figure 9.** A comparative view on the $sp^2$-$sp^3$ transformation of the $C_{60}$ cage structure of 18-fold derivatives under fluorination, hydrogenation, cyanation, and aziridination.

As it turns out, a stepwise synthesis of chlorides is subordinated to a sequence of 1,2-additions as in all cases considered earlier. Thus, a predicted preference of first-step 1,4-additions of chlorine atom to the $C_{60}$ cage over 1,2-ones [16], which is rather small, is inverted when the distribution of effectively unpaired electrons over cage atoms is taken into account. However, the chlorination proceeds absolutely in a different way in regards to, say, fluorination. This difference is originated at the third step of the reaction when the third chlorine atom is attached to the cage after the first two were situated at two carbon atoms belonging to group 1. In contrast to the third fluorine (hydrogen, cyan) and the second NH units, which were to be settled in the equatorial space of the fullerene cage, the third chlorine atom takes a seat in the nearest vicinity of the first two atoms revealing a tendency to a contiguous sequence of the addition steps. It should be noted that the preference of the highest rank $N_{DA}$ data over remaining ones is the strongest in the chlorination case in regards to four processes considered earlier.

The contiguous addition occurs until chloride $C_{60}Cl_6$ is formed. Figure 10 presents $C_{60}F_6$ (Fig. 10*a*) and $C_{60}Cl_6$ (Fig. 10*b*) structures to be compared. A non-contiguous and contiguous sequence of the addition steps is clearly seen. The contiguity of chlorine addition is terminated at the sixth step and subsequent 1,2-additions occur quite far from the first six atoms (see Fig. 6.10), leading to an evidently symmetric structure of chloride $C_{60}Cl_{18}$, shown in Fig. 11. In spite of absolutely different pattern of the structure in comparison with that of $C_{60}F_{18}$, the added atoms are concentrated in the upper part of the molecules form a crown-like 'flower-bowl' in the first case and 'flower-bowl with arms' in the second. The addition of the twentieth chlorine atom causes a destruction of the cage that proceeds when the number of the attached atom increased. The chemical bond rupture greatly raises the cage radicalization, which, in its turn, strengthens chemical reactivity of the chlorides stimulating their further chlorination.

In 1991 was shown that fullerene $C_{60}$ easily reacted with both gaseous and liquid chlorine and originated a mixture of species with either 12 or 24 chlorine atoms in general per one molecule of fullerene [15, 17]. This appears to be connected with the fullerene cage destruction due to massive chlorination mentioned above. The first individual chloride $C_{60}Cl_6$ was obtained two years later [18] and until now it has been the most studied representative of the family [19, 20]. The chloride structure was attributed to $C_s$ symmetry that is consistent with that of $C_{60}Cl_6$ species given in Fig. 10*b* (chloride *A*). However, the structure of the $C_{60}Cl_6$ molecule presented in Fig. 10*c* (chloride *B*), which was proposed on the basis of numerous synthetic data [19, 20] where the chloride was used as initial chemical reactant, differs from the calculated quite drastically. Since the energy of molecule *B* is much lower than that of molecule *A*, and since chlorine atoms can easily move over the fullerene surface [21], one may suggest that initially formed *A* structure is transformed into *B* one due to 'chlorine atom dancing', simultaneously causing a reizomerization of the fullerene cage to suit new requirements concerning pentagon framing similarly to the case of $C_{60}F_{48}$ fluoride discussed in [1].

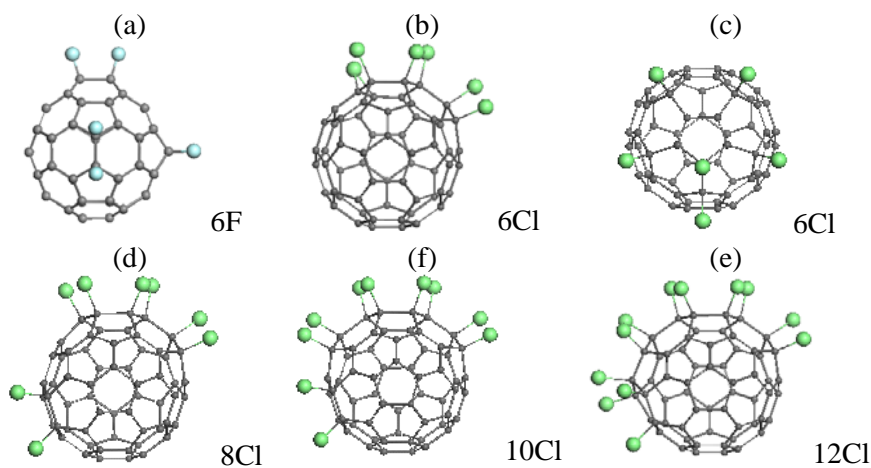

**Figure 10.** Equilibrium structures of fluoride $C_{60}F_6$ (*a*) and chlorides $C_{60}Cl_6$ (A) (*b*), $C_{60}Cl_8$ (*d*), $C_{60}Cl_{10}$ (*f*), and $C_{60}Cl_{12}$ (*e*). UBS HF AM1 singlet state. Chloride $C_{60}Cl_6$ (B) suggested on the basis of experimental data [19, 20] (*c*). Small black balls point carbon atoms while light blue and green balls mark fluorine and chlorine atoms, respectively

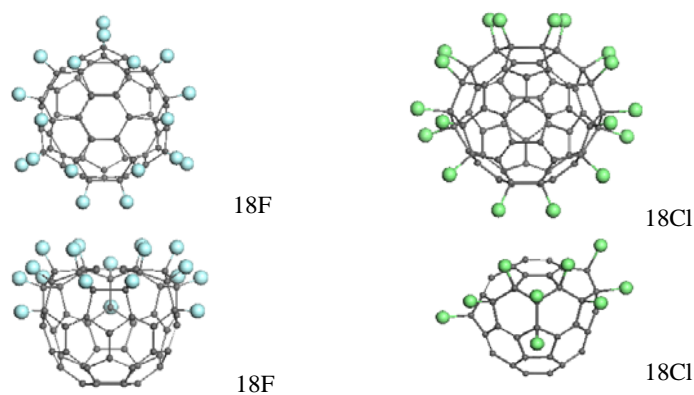

**Figure 11.** Two projections of the equilibrium structures of fluoride $C_{60}F_{18}$ and chloride $C_{60}Cl_{18}$. UBS HF AM1 singlet state. Atoms marking see in caption to Fig. 10.

Besides $C_{60}Cl_6$, other chlorides such as $C_{60}Cl_8$, $C_{60}Cl_{12}$, $C_{60}Cl_{28}$, and $C_{60}Cl_{30}$ were produced under severe experimental conditions [13]. As seen from the list, it differs drastically from those of fluorides and hydrides and does not involve 18-, 24-, 36-, and 48-fold species thus supporting a cardinal difference of the chlorination process. A contiguous-addition initial stage of the chlorination seems to be responsible for the difference. Therefore, to specify the structure evolution of a derivative family, three criteria involving sterical limitations, 1,2-manner of additions and a sequence of the additions on the first stage of the reaction should be taken into account.